\documentclass[11point]{article}
\begin{document}
\title{\Large  \textbf{Role of Topologies in the Study of Dynamical Aspects of Stable and Generic Properties of Space-times in General Relativity}}
\date{}
\maketitle{\noindent\small {{R \ V \ Saraykar $\ ^{1}$ and  Sujatha \ Janardhan$\ ^{2}$}}  \\$\ ^{1}$ {Department of Mathematics,
R\ T\ M Nagpur University, Nagpur-440033. \\ e-mail :  ravindra.saraykar@gmail.com
\\ $\ ^{2}$ {Department of Mathematics, St.Francis De Sales College,  Nagpur-440006. \\ e-mail :
sujata\_jana@yahoo.com}}}

\vspace{10mm}

\noindent\textbf{Abstract}: In [1], the authors have studied stability of certain causal properties of space-times in general relativity. As a continuation of this work, in the present paper, we review and discuss, some more aspects of stability which occur in various situations in the dynamics of general relativity. We argue that not only choice of appropriate topology, but also the nature of perturbation, like metric, matter or initial data, are key factors in deciding whether a property is stable or not. We also study certain properties of space-times which are generic in an appropriate mathematical sense. In particular we argue that Linearization stability of a space-time is a generic property.
\section{Introduction}

A space-time is a four dimensional differentiable manifold $V$ endowed with a Lorentz metric. Let $Lor(V)$ denote the set of all Lorentz metrics which can be defined on $V$. A causal property or a global property $P$ of a space-time is said to be stable if the set of Lorentz metrics possessing the property $P$ forms an open subset of $Lor(V)$. Since this definition involves open sets, we need to define a topology on $Lor(V)$. Since there are various ways of defining topology on $Lor(V)$, there is no unique definition of stability. Various such topologies and concept of stabiity for various causal properties have been described in details in [1] and references therein. We shall not go into the details of this treatment here, and we refer the reader to [1] for details in various cases. Moreover, we would like to prove that a property holds for all members of $Lor(V)$, or the property holds at least for 'almost all' members of $Lor(V)$.  This means the theorem may hold for almost all metrics in the region, but fail for some particular metrics. In this situation, we say that such a theorem holds generically. In general, a property is said to be generic in a region of $Lor(V)$ if it holds almost everywhere on that region.
Mathematically, by 'almost everywhere' we mean that it holds on an open dense subset of the region of $Lor(V)$. Similar definition is given in the theory of Dynamical Systems also. ( See for example, Abraham and Marsden [2]). 
From the definitions of "Stability" and "Genericity", it is clear that these concepts depend upon
the topologies under consideration. Thus a given property may be stable and generic in some
topologies and not in others. Which of the topologies is of physical interest will depend upon the
nature of the property under consideration and also the nature of perturbation that we are considering. Perturbation can be a metric perturbation, or change in the nature of matter in a space-time or slight change in the initial data from which a space-time evolves. 
Thus, in this article, we study some more aspects of stability excluding those in [1]. Furthermore, we note that, while studying stability of global properties (Cf.[1]), we did not take into consideration the Einstein field equations which are evolution equations, and hence dynamical aspects were not taken into account. Here, we consider these aspects. In particular, we discuss linearization stability of Einstein field equations and some important results proved in the past. We also consider briefly certain examples from the study of dynamical gravitational collapse to emphasize that nature of outcome of collapse changes if matter in the space-time changes. This is another aspect of stability.
Furthermore, we discuss the issue of genericity which occurs in four different contexts. 

Thus, in Section 2, we study various aspects of stability as mentioned above. We also discuss an example of gravitational collapse. In Section 3, we discuss properties of space-times occurring in four different contexts which are generic in the above sense. In particular, we argue that Linearization Stability is a generic property.  At the end we make concluding remarks on both these important issues of stability and genericity.\\

\section{Dynamical aspects of Stability}

\textbf{Linearization Stability :}
For the sake of completeness and convenience of the reader, we borrow following definition and some properties from Fischer and Marsden [3] :\\

Definition : Let $\Phi: X \longrightarrow Y$ be a non-linear differential operator between Banach spaces or Banach manifolds of maps $X$ and $Y$. Consider the equation $\Phi(x)$ = $y_{0}$ for $y_{0} \in  Y$.

Let $ T_x{X} $ denote the tangent space to $ X $ at $x \in X$, and let \\ $D\Phi(x)$ : $T_x{X} \longrightarrow T_y{Y}$, with $y$ = $\Phi(x)$, be the Frechet derivative of $\Phi(x)$ at $x$. Thus to each solution $x_0$  of $D\Phi(x)$ = $y_0$ ,  $D\Phi(x_0).h = 0$, $h \in T_{x_{0}}X$, is the associated system of linearized equations about $x_0$, and a solution $h \in T_{x_0}X$ of linearized equations is an infinitesimal deformation ( or first order deformation ) of the solution $x_0$.
If for each solution $h$ of linearized equations,there exists a curve $x_t$   of exact solutions of $D\Phi(x) = y_0$  which is tangent to $h$ at $x_0$  i.e. $x(0) = x_0$  and $[(d / dt)(x_t )]_{|_{t = 0}} = h$, then we say that equation $\Phi(x) = y_{0}$ is linearization stable at $x_0$, and deformation $h$ is called integrable.

Useful criterion to prove linearization stability is as follows :\\
Theorem: Let $X$ and $Y$ be Banach manifolds, and $\Phi: X \longrightarrow Y$ be a $C^1$-map. Let $x_0 \in X$ be a solution of $\Phi(x) = y_{0}$. Suppose $D\Phi(x_0)$ is surjective with splitting kernel . Then the equation $\Phi(x)$ = $y_{0}$ is linearization stable at $x_0$.\\
Here, splitting kernel means:  $T_{x}X = Range (D\Phi)^*(x) + Ker D\Phi(x)$). \\
Proof uses the Implicit Function Theorem.\\
This definition applied to Einstein field equations in the case where Cauchy (spacelike) hypersurface is a compact 3- manifold $M$ without boundary can be phrased as follows :\\

We write Einstein equations for vacuum space-time as : $Ein(^4{g}) = 0$, where Ein denotes Einstein tensor. Let $^4{g}_0$ be a solution of Einstein equation. Then linearized equation is given by $DEin(^4{g}_0).^4{h}  = 0$. If for every $^4{h}$ satisfying linearized Einstein equation, there is a curve ($^4{g}_ {t}$) of exact solutions such that $[^4{g}_t]_{|t = 0}  = ^4{g_0}$ , and $[(d / dt)(^4{g_t} )]_{|_{t = 0}} =  ^4h$, then $^4{g}_0$ is called linearized stable.  In this case $^4{ h}$  is called integrable.

Brill and Deser [4] were the first to show that not every $^4{ h}$ satisfying Linearized equation is integrable. They considered space-time $({T^3})\times R$ and showed that there are solutions of linearized equation which are not integrable.
In the ADM formalism for Einstein field equations, we can split these equations into six evolution equations and four constraint equations. In the case when a space-time admits a compact Cauchy hypersurface $M$, all dynamics of space-time is absorbed in the constraint equations. This is because Cauchy problem is well-posed. Moreover, it is also well-known that linearization stability of Einstein field equations is equivalent to that of constraint equations. If we denote constraint equations by 
$\Phi(g,\pi)=0$, where $g$ denotes three dimensional Riemannian metric on Cauchy hypersurface $M$ and $\pi$ is the corresponding conjugate momenta, then we have the following results, which are proved in [3]:\\ 

(1) For space-time $V$ admitting a compact constant mean curvature (CMC) space-like hypersurface, $(D\Phi)^*(g,\pi)$ is elliptic. \\
(2) Let space-time $(V, {^4{g}})$ be fixed. The space of Killing fields of $^4{g}$  is isomorphic to the kernel of $(D\Phi)^*(g,\pi)$. Here, we note that an elliptic operator has a finite dimensional kernel.\\
(3) If $(V, {^4{g}})$  has no Killing fields, then it is linearization stable.\\

Thus, if $(V, {^4{g}})$ is linearization stable , ker $(D\Phi)^*(g,\pi)$ is trivial.
Since the space-time $({T^3})\times R$ possesses symmetries (Killing fields), it is now clear from above results as to why Brill-Deser[4] could find non-integrable perturbations.\\

Thus, space-times admitting compact CMC hypersurfaces can admit only finite number of independent Killing fields. Similar results hold for Einstein field equations coupled with matter fields such as scalar fields, electro-magnetic fields and Yang-Mills fields. See, for example, $[5,6,7]$. To analyse the situation in the presence of Killing fields, we need techniques from non-linear functional analysis. We first identify a four dimensional Killing field $^4X$ with a pair $(N,X)$, where $N$ is the perpendicular component of $^4X$ and $X$ is its parallel (space-like) component. $N$ and $X$ are a real-valued function and a vector field on $M$ respectively. In ADM-formalism, $N$ and $X$ play the role of a lapse function and shift vector field. Thus, if $^4X$ (equivalently $(N,X)$) is a Killing field, then integrable Linearized perturbations $(h,w)$ of $(g,\pi)$
satisfy a second order condition :-

$ \int_M <(N,X), (D^2\Phi(g,\pi)\mid\cdot ( (h,w), (h,w) ) )>  = 0 $ \\
It is also true that this condition is sufficient for a linearized perturbation to be integrable. The idea of the proof of this non-trivial statement is as follows :\\
We need to show that the above second order condition is non-vacuous. i.e. if $^4{X}\neq 0$ is a Killing field, then there exists $(h,w)$ satisfying the linearized equations such that above second order quantity does not vanish. This implies that $(h,w)$ is not integrable
and hence $(V, {^4g})$ is not linearization stable. The hypersurface-invariance of second order
quantities, explicit expression of $D^2\Phi(g,\pi)$ and elliptic character of $D\Phi(g,\pi)^*$ play
an important role in the proof.
The main result of this analysis is described as follows :\\
Consider a class $\textbf{\textsl{E}}$ of space-times represented by Lorentz metrics $^4g$ and let $(V,{^4g_0})$ be a space-time admitting a compact cauchy hypersurface of constant mean curvature. Suppose $(V,{^4g_0})$ admits $n$ independent Killing fields. Then the space $\textbf{\textsl{E}}$ has a conical singularity at $^4g_0$.
This means, in the neighbourhood of $^4g_0$, $\textbf{\textsl{E}}$ can be written in a suitable chart as the zero set of a
homogeneous quadratic function. The generators of this cone consist of those symmetric
two-tensors $^4h \leftrightarrow (h,w)$  such that

(i)\ $^4h$ satisfies the Linearized Einstein equation
$D Ein(^4g_0).{^4h} = 0$ or equivalently, $D\Phi(g_0,\pi_0).(h,w) = 0$,\\
and (ii) the second order condition above is satisfied.\\
Thus, a necessary and sufficient condition for a solution $^4h$ of the linearized equations
to be tangent to a curve of exact solutions to Einstein equations passing through $^4g_0$ is that above
second order condition is satisfied.
The proof uses involved mathematical machinery like slice theorem for action of a Lie
group on a manifold, Kuranishi map from deformation theory of complex manifolds and
infinite dimensional Morse lemma due to A. Tromba. For technical details of this work, we refer the reader to Fischer, Marsden and Moncrief [8], Arms, Marsden and Moncrief [9] and Saraykar [10].\\

\textbf{Instability of global hyperbolicity through metric and matter perturbation}:
 Another way in which we can discuss stability issue is with respect to matter as well as metric perturbation. Consider, for example, the Schwarzschild space-time. It is globally hyperbolic. But if we throw a smallest charge into it, then the resulting space-time is Reissner-Nordstrom, which is not globally hyperbolic. Moreover it admits a Cauchy horizon, and Cauchy surfaces are lost(Cf. Hawking and Ellis [11]). As another example, consider the Oppenheimer-Snyder collapse, i.e. the spherically symmetric dust collapse, which is globally hyperbolic. But with a slightest perturbation in density, e.g. the density higher at the center, we loose global hyperbolicity, and the singularity, which was earlier a black hole, now  becomes a naked singularity. Now choosing a slightly higher density at center corresponds to a certain perturbation of the metric. Thus, it will be interesting to consider such a class, which is of course physically interesting, and  try and see why global hyperbolicity is violated ? So should we call it stable or not? In the same example, if density is inhomogeneous and if we add slight pressure in the collapsing system, then again the outcome can change from black hole to a naked singularity or vice-versa. Thus the outcome can be called unstable with respect to addition of pressure.\\
\textbf{Stability of end state of gravitational collapse with respect to initial data}:
End state of gravitational collapse of type I matter fields evolving from regular initial data can be either a black hole or a naked singularity. We ask the question : Is this outcome stable with respect to slight change in initial data ? Saraykar and Ghate [12] answered this question in the affirmative in the case of inhomogeneous dust collapse. Furthermore, in the case of type I matter fields, this question was settled in a positive matter by Sarwe and Saraykar [13], and all mathematical details were provided by Joshi, Malafarina and Saraykar [14]. In these works, it was proved  that the outcome is a stable one, but not generic, in the sense that the subset of initial data which leads the collapse to a black hole or a naked singularity forms an open subset of the set of all initial data with an appropriate topology. The topology used here is the $C^1$ topology on the set of functions. However, this set is not dense. But, using the measure on infinite dimensional spaces, the authors of [14] could prove that the initial data set leading the collapse to a particular outcome has non-zero measure. For review on this topic and other aspects of stability as mentioned above, we refer the reader to Joshi and Malafarina [15]. It is clear that in this work also, topology on a function space plays an important role in deciding the stability.\\

\textbf{Stability with respect to odd and even parity perturbations }\\
Another way of investigating stability of space-times is by considering odd and even parity perturbations of the space-time metric and see if these perturbations are integrable. There is a vast literature on this aspect, and we shall not go into the details of this area. Instead, we refer the reader to an excellent book by S. Chandrasekhar [16].\\

\section{Generic properties occurring in general theory of relativity}

As mentioned above, a generic property is one that holds on a dense open set, or more generally on a residual set (a countable intersection of dense open sets).   The spaces that are usually used to study stability and genericity are function spaces whose elements are either vector fields or tensor fields which are continuous mappings with $r$ continuous derivatives from a certain manifold $M$ to a manifold $N$. We denote such function spaces by $C^r(M,N)$. A property is said to be generic in $C^r(M,N)$ if the set holding this property contains a residual subset in the $C^r$ topology. This topology is often used in the mathematics literature, and is called Whitney-$C^r$ topology.  We also note that the space $C^r(M,N)$ of $C^r$ mappings between $M$ and $N$ is a Baire space and hence any residual set is dense. This property of the function space is what makes generic properties typical. The generic properties that we study in this section are as follows :\\
 $(i)$ A property of a space-time being stably causal is a generic property.\\
 $(ii)$ Linearization stability is a generic property in an appropriate function space. Specifically, Saraykar and Rai [17] proved that in the class $\mathcal{V}$ of space-times which admit a compact Cauchy hypersurface of constant mean curvature, the subclass $\mathcal{V}_K$ of space-times which are linearization stable form an open and dense subset of $\mathcal{V}$ under $C^\infty$ - topology.\\
  $(iii)$ The "generic condition" occuring in the Hawking-Penrose singularity theorems (Cf.[11]) is generic. Beem and Harris [18] proved that the set of vector fields satisfying this condition is open and dense in the set of all vector fields under a suitable topology.\\
  $(iv)$ Fourth instance of genericity is the theorems proved by Ringstrom [19,20] in the quest of proving the strong cosmic censorship conjecture. He proves that under suitably defined topology, the set of initial data evolving into Einstein vacuum equations is open and dense in the set of all initial data. This has been proved for (${T^3}\times{R}$)-Gowdy space-times.\\   
We now discuss these properties one by one.\\
\textbf{1. Stable causality is a generic property}\\
 The region in $Lor(V)$ in which stable causality holds lies in the interior of the region on which ordinary causality holds. Also the region in which ordinary causality is violated, is open in $C^0$ open topology. Hence the union of this region with the region on which stable causality holds is an open dense set in $Lor(V)$. It is thus generic for a metric either to violate causality or to be stably causal.
Hawking [21] conjectures that the subset of stably causal metrics is dense in the set of all causal metrics.\\
\textbf{2. Linearization Stability is a generic property }:\\
Here we consider the class $\mathcal V$ of all space-times ( equivalently class of Lorentz metrics ) possessing compact Cauchy hypersurfaces of constant mean curvature, and  use results of Beig, Chrusciel and Schoen [22] to argue that the subclass ${\mathcal V}_K$  of $\mathcal V$ possessing no Killing fields forms an open and dense subset of $\mathcal V$.
Combining this result with above results for linearization stability discussed in Section $2$, we conclude that the class of space-times possessing compact Cauchy hypersurfaces of constant mean curvature which are linearization stable forms an open and dense subset of $\mathcal V$. Thus, in this sense, linearization stability is a generic property.\\
\textbf{Outline of the Proof :}\ We consider the class $\mathcal V$ of all vacuum space-times as above. Such a class of space-times can be endowed with a suitable topology whose choice can be made as per our need. Such topologies have been discussed by Hawking $[21]$, Lerner $[23]$ and Beig, Chrusciel and Schoen $[22]$. Then we use the results on linearization stability as follows.\\
As discussed in Section $2$, space-times admitting compact CMC hypersurfaces can admit only finite number of independent Killing fields. Also, such a space-time $(V, {^4{g}})$ is linearization stable if and only if it has no Killing fields if and only if ker $(D\Phi)^*(g,\pi)$ is trivial. We also have the fact that any vector field $^4X$ on a space-time can be identified with a pair $(N,X)$ as explained earlier. This decomposition is of course related to Riemannin metric $g$ on the Cauchy hypersurface $M$. Beig et.al [22] call this pair $(N,X)$ as Killing Initial Data (KID) when $^4X$ is a Killing field on $(V,{^4{g}})$, and $(g,\pi)$ is called vacuum initial data, where $(g,\pi)$ is as in ADM formalism as explained above. Thus, each space-time $(V,{^4{g}})$ corresponds to a vacuum initial data $(g,\pi)$ and the Killing field $^4X$ which a space-time admits, corresponds to a KID $(N,X)$. Due to covariance property of Einstein field equations, Killing property of initial data will be carried throughout the evolution. Thus this property is possessed by space-time as a whole. In other words, KIDs are in one-to-one correspondence with Killing vectors in the associated space-time. With this association, and keeping above result in mind, it is now clear that the class of vacuum space-times $\mathcal{V}_K$ possessing a compact cauchy hypersurface which are linearization stable can be identified with the class $\mathcal{V}_I$ of all vacuum initial data $(g,\pi)$ without (global) KID with $tr(\pi)$ = constant. To this class $\mathcal{V}_I$, we apply a theorem proved in [22] and conclude genericity result as mentioned above. For more details, we refer the reader to [17].\\

Finally, in the context of general relativity, it is interesting to note that it is not yet fully known if the above generic result is valid when constant mean curvature condition is removed.\\

\textbf{3. Generic condition is generic} : \\
Let $(V,{^4{g}})$  denote a space-time and $R_{ab}$  denote Ricci tensor corresponding to the Lorentz metric $^4g$. Let $\gamma(v)$ denote a timelike or null curve in $V$ and let $p = \gamma(v_1)$. Let $K$ denote a general tangent vector. Then the following result holds (Cf. Hawking-Ellis [11], Prop. 4.4.5, page 101) :

Result 1 : If $R_{ab}K^{a} K^{b} \geq 0$ holds everywhere and if at $p = \gamma(v_1)$, $K^{c} K^{d} K_{[a}R_{b]cd[e}K_{f]}$ is non-zero, then there will be $v_0$ and $v_2$ such that $q=\gamma(v_0)$ and $r=\gamma(v_2)$ will be conjugate along $\gamma(v)$ provided $\gamma(v)$ can be extended to these values.

Hawking-Ellis explain that it is reasonable to assume that in a physically realistic space-time, every timelike or null geodesic will contain a point at which the quantity $K^{c} K^{d} K_{[a}R_{b]cd[e}K_{f]}$ is non-zero. This condition is called 'generic condition'. It can be satisfied by a single tangent vector, or by every tangent vector to a null or timelike curve. In the latter case, we say that a space-time itself satisfies the generic condition. It is well-known [11] that the generic condition for a space-time plays an important role in the proof of singularity theorems. Beem and Harris[18$(V,{^4{g}})$ ] prove that this Generic condition is generic in the sense that at a given point of spacetime, if we consider the tangent space at this point, then the set of tangent vectors satisfying the generic condition is open and dense in the tangent space. Beem and Harris prove a series of results which are algebraic in nature, and then the above result follows as a consequence of these. We describe these results in brief :

A tangent vector $K$ is called non-generic if it is non-zero and if $K^{c} K^{d} K_{[a}R_{b]cd[e}K_{f]}$ = $0$. Thus, a causal geodesic satisfies the generic condition if and only if its velocity vector is not everywhere non-generic.

Following series of results have been proved in [18] : \\
Result 2 : The set of non-generic vectors, together with the zero vector, is a closed set. Furthermore, any null vector which is a limit of non-generic non-null vectors or of strongly non-generic null vectors is itself strongly non-generic. \\
Hence the set of generic vectors is an open set.\\
Result 3 : The entire vector space $V$ is non-generic if and only if it is flat.\\
Result 4 : Let $V$ be four dimensional. Suppose $V$ has $5$ non-null non-generic vectors, with $4$ forming a basis and the fifth not in the plane spanned by any two of those basis vectors. Then $V$ is flat.\\
Result 5 : Let $W$ be a subspace of $V$ with codimension of $W$ = $1$. If $W$ is non-generic, then $V$ is flat.\\
Result 6 : Let $W$ be a subspace of $V$ with dim($W$)= $m$. Suppose $W$ has the following non-null non-generic vectors $\left\{X_i : {1 \leq i \leq m}\right\}$, which span $W$; and for each $i<j$, $Y_{ij}$ in span $({X_i, X_j})$, such that $(X_i, X_j,Y_{ij})$ is in general position for span $({X_i, X_j})$. Then $W$ is non-generic.\\
Applying this result to dim($V$) = $4$, and dim ($W$) = $3$, we get the following corollary of result $6$ :\\
Result 7 : If $V$ has a subspace $W$ of codimension 1, so that $W$ contains a set of non-generic vectors which is open in $W$, then $V$ is flat.

\textbf{Theorem:} Generic condition is generic, in the sense that at a given point of spacetime, if we consider the tangent space at this point, then the set of tangent vectors satisfying the generic condition is open and dense in the tangent space.\\
\textbf{Proof:} Denseness property of generic vectors now follows from this corollary (Result 7) as argued below :\\
If the generic vectors do not form a dense set, then there is an open set of non-generic vectors, whose intersection with any subspace is open. This implies, by above corollary, that $V$ is flat; which gives a contradiction. Thus, generic vectors form a dense set. That it forms an open set, follows from Result 2 mentioned above. Thus, set of generic vectors forms an open and dense subset of the set of all tangent vectors at a given point, i.e.the tangent space at that point. This then proves that generic condition applied to a tangent vector is really a generic property.\\
Globally i.e. considering the space $Lor(V)$, it can be proved that generic condition is satisfied for all metrics
in a residual set in the Whitney-$C^r$ topology, where $r$ depends upon the dimension of the manifold $V$.\\
Our last example of genericity is from the work of Ringstrom where he proves strong cosmic censorship conjecture for a certain class of space-times.\\
4. \textbf{Set of initial data (Cauchy surface ) from which a space-time develops, in the sense of maximally globally hyperbolic development, is open and dense in the class of all initial data in a suitable topological sense	}( Ringstrom [19,20] ) :

Einstein's vacuum equations can be viewed as an initial value problem, and given initial data there is one part of space-time, the so-called maximal globally hyperbolic development (MGHD), which is uniquely determined up to isometry. However, it is sometimes possible to extend the space-time beyond the MGHD in inequivalent ways. Hence, the initial data do not uniquely determine the space-time, and in this sense the theory is not deterministic. Here, it is then natural to make the strong cosmic censorship conjecture, which states that for generic initial data, the MGHD is inextendible. Since it is unrealistic to hope to prove this conjecture in all generality, it is natural to make the same conjecture within a class of space-times satisfying some symmetry condition. Ringstrom, in a series of two papers, proved strong cosmic censorship in the class of (${T^3}\times{R}$)-Gowdy spacetimes.

In the first paper, the author focuses on the concept of asymptotic velocity. Under the symmetry assumptions, Einstein's equations reduce to a wave map equation with a constraint. The range of the wave map is the hyperbolic plane. The author introduces a natural concept of kinetic and potential energy density. The important result of this paper is that
the limit of the potential energy as one lets time tend to the singularity for a fixed spatial point is $0$ and that the limit exists for the kinetic energy.

In the second paper [20], the author proves that the set of initial data $G_i$ is open with respect to the $\ C^1$ topology and dense with respect to the $\ C^\infty$ topology, such that the corresponding space-times have the following properties:

First, the MGHD is $C^2$-inextendible. Second, following a causal geodesic in a given time direction, it is either complete, or a curvature invariant, the Kretschmann scalar is unbounded along it (in fact the Kretschmann scalar is unbounded along any causal curve that ends on the singularity).\\
This completes our study of generic properties occuring in general theory of relativity.\\

\textbf{Concluding Remarks :}

 From our studies and discussion in sections 2 and 3, it is clear that when the definitions of stability and genericity involve openness of the data, then the choice of topology plays an important role in deciding these properties. Thus, we can not expect uniqueness in the choice of definition of stability and genericity. Just as this is true for problems in general theory of relativity, it is also true in other branches of pure and applied mathematics. \\

\textbf{Acknowledgdement :} The author wishes to express sincere thanks to Prof. Pankaj S. Joshi, Tata Institute of Fundamental Research, Mumbai, for many helpful comments, suggestions and discussions about this work.

 \section{References}

  1. R. V. Saraykar and Sujatha Janardhan, Topologies on the Space of Lorentz Metrics and Stability of Global Properties of a Space time Manifold, Journal of the       Tensor Society,Vol. 6(2), 95-106, (2012).\\
  2. R. Abraham and J.E. Marsden, Foundations of Mechanics, 2nd edition, Addison-Wesley (1978).\\
  3. A.E. Fischer A. and J.E. Marsden, Topics in the dynamics of general relativity, in 'Isolated gravitating systems in general relativity', Ed. J. Ehlers,            Italian Physical Society, (1979), 322-395.\\
  4. D. Brill and S. Deser, Instability of closed spaces in general relativity, Commu. Math. Phys. Vol. 32, 291-304 (1973) \\
  5. R.V. Saraykar and N.E. Joshi, Linearisation stability of Einsteins equations coupled with self-gravitating scalar fields,Jour. Math. Phys. Vol. 22,                343-347(1981);Erratum Vol.23, (1982), 1738. \\
  6. J. Arms, Linearization stability of Einstein-Maxwell fields, Jour. Math. Phys.Vol.18, 830-833(1977).\\
  7. J. Arms, Linearization stability of gravitational and gauge fields, Jour. Math. Phys.Vol. 20, 443-453(1979).\\
  8. A.E. Fischer, J.E. Marsden and V. Moncrief, The structure of the space of solutions of Einstein's equations.I. One Kiling field,Annales de la Institut H.          Poincare, Section A, Vol.33 (2) (1980), 147-194.\\
  9. J. Arms, J.E. Marsden and V. Moncrief, The structure of the space of solutions of Einstein?s equationsII. Several Kiling fields and the Einstein-Yang-Mills        Equations, Annals of Physics, Vol. 144 (1) (1982), 81-106.\\
  10.R.V. Saraykar, The structure of the space of solutions of Einsteins equations coupled with Scalar fields, Pramana, a Journal of Physics, Vol.20,no.4, 293 -          303 (1983).\\
  11. S.W. Hawking and G.F.R. Ellis, The large scale structure of space-time, Cambridge University Press, (1973).\\
  12. R.V. Saraykar and S.H. Ghate, $C^1$-stability of naked singularitites arising in an inhomogeneous dust collapse, Class. and Quantum Grav., Vol. 16, 281-291 (1999).\\
  13. S.B. Sarwe and R.V. Saraykar, Stability of naked singularity arising in gravitational
Collapse of type I matter fields, Pramana, Jour. of Phys. ,Vol.65, no.1, 17-33 (2005). \\  
  14. Pankaj S. Joshi, Daniele Malafarina and Ravindra V. Saraykar, Genericity aspects in gravitational collapse to black holes and naked singularities, Inter.          Jour. Mod. Phys. D, Vol. 21, No.8  ( 2012) , (38 pages )\\
  15. Pankaj S. Joshi and Daniele Malafarina, Recent developments in gravitational collapse and space-time singularities, Inter. Jour. Mod. Phys. D, Vol. 20,           2641 (2011) , (89 pages ).\\
  16. S. Chandrasekhar, Mathematical theory of black holes, International Series of Monographs on Physics, Vol. 69. Clarendon Press Oxford, 1983.\\
  17. R.V. Saraykar and Juhi H. Rai, LinearizationStability of Einstein Field Equation is a Generic Property, Electronic Jour. Theor.Phys.Vol. 13, 36, 1-10,             (2016).\\  
  18. J.K. Beem and S.G. Harris, Generic condition is generic, Gen. Rel. and Grav. Vol. 25(9),939-961(1993).\\
  19. H. Ringstrom, Existence of an asymptotic velocity and implications for the asymptotic behavior in the direction of the singularity in T3-Gowdy, Commun. in          Pure and Appl. Math.Vol.59, 977-1041,(2006).\\
  20.. H. Ringstrom, Strong cosmic censorship in $T^3$-Gowdy spacetimes, Annals of Math., Vol. 170, 1181-1240, (2009).\\
  21. S.W. Hawking, Stable and generic properties in general relativity, Gen. Rel. Grav. Vol. 1, 393-400 (1971).\\
  22. Robert Beig, Piotr T. Chrusciel and Richard Schoen, KIDs are non-generic, Ann. Inst. H. Poincare, Vol. 6, 155-194 (2005) \\
  23. D. Lerner, Topology on the space of Lorentz metrics, Commu. Math. Phys., Vol.32 , 19 - 38, (1973).\\

\end{document}